\documentstyle[12pt,twoside,fleqn,espcrc1,psfig]{article}

\newcommand{\AmS}{{\protect\the\textfont2
  A\kern-.1667em\lower.5ex\hbox{M}\kern-.125emS}}
\title{Measurement of direct neutron capture by neutron--rich sulfur isotopes}

\author{H. Beer\address{Forschungszentrum Karlsruhe, IK III,
                P.O.~Box 3640, D--76021 Karlsruhe, Germany}, %
                C. Coceva\address{ENEA, Via Don Fiammelli 2, I--40128 Bologna, Italy}, %
                R. Hofinger\address{Institut f\"ur Kernphysik, TU Wien,
                Wiedner Hauptstra{\ss}e 8--10, A--1040 Wien, Austria}, %
                P. Mohr$^{\rm c}$, %
                H. Oberhummer$^{\rm c}$, %
                P.V. Sedyshev\address{Frank Laboratory, JINR,
                141980--Dubna, Moscow Region, Russia} %
                and %
                Yu.P. Popov$^{\rm d}$}
                
\begin{document}
\maketitle

\begin{abstract}
Thermal neutron capture cross sections for $^{34}$S(n,$\gamma$)$^{35}$S
and $^{36}$S(n,$\gamma$)$^{37}$S have been measured and
spectroscopic factors of the final states have been extracted. The calculated
direct--capture cross sections reproduce the experimental data.
\end{abstract}

\section{INTRODUCTION}

Reaction rates of the neutron--rich S--isotopes are of
interest in the nucleosynthesis of nuclei in the s--process
in the S--Cl--Ar--Ca region,
inhomogeneous big--bang scenario,
and in the $\alpha$--rich freeze out of the
neutron--rich hot neutrino bubble in supernovae of
type II. In the last years the importance of the
direct capture (DC) has been realized.
DC dominates over the
compound--nucleus reaction mechanism if there exist no
compound--nucleus levels near the threshold that can be excited in the
reaction. This is often the case for neutron capture for neutron--rich target
nuclei at the border of stability.
%
%
With respect to astrophysical applications thermal measurements of such
capture reactions are an important supplement to measurements in the
thermonuclear energy region. Such measurements allow a more reliable
extrapolation to lower energies. Furthermore, when there are no
resonant contributions at thermal energy, experimental thermal cross
sections allow the extraction of spectroscopic factors. These
spectroscopic factors are needed to calculate the DC contribution
at thermonuclear energies. This extraction is not as much model--dependent
as the extraction of spectroscopic factors from direct
transfer reactions, like (d,p), ($^3$He,$^4$He), or other
transfer reactions at higher energies (about 15--40\,MeV). This is mainly
due to the fact that the neutron optical potentials necessary for
the extraction of the spectroscopic information from
experimental thermal (n,$\gamma$)--data is better determined than the
optical proton and especially light-ion
potentials at higher energies (because for higher energies
there is a larger absorption).

The purpose of this work is to show that for thermal
neutron capture by target isotopes close
to the border of stability, like $^{34}$S and $^{36}$S, the DC mechanism
\begin{table}[hbt]
\caption{The intensity per incident neutron,
$I_{\gamma}$, and statistical uncertainty, $\Delta I_{\gamma}$, of the 
$\gamma$--ray transitions with initial and final spin/parity,
$J_{\rm i}^{\Pi}$ and $J_{\rm f}^{\Pi}$, respectively, and the transition 
energy $E_{\gamma}$. The partial and total capture cross
sections of $^{34,36}$S, $\sigma^{\rm p}$ and $\sigma$, were determined.} 
\label{t2a}
\begin{tabular*}{\textwidth}{@{}l@{\extracolsep{\fill}}rrrrrr}
\hline
\multicolumn{1}{c}{Target} &
\multicolumn{1}{c}{E$_{\gamma}$$^{\rm a}$} &
\multicolumn{1}{c}{Transition} & 
\multicolumn{1}{c}{I$_{\gamma}$$\pm$$\Delta$I$_{\gamma}$} &
\multicolumn{1}{c}{$\sigma^{\rm p}$} &
\multicolumn{1}{c}{$\sigma$}\\
\multicolumn{1}{c}{Isotope} &
\multicolumn{1}{c}{(keV)} &
\multicolumn{1}{c}{J$_{\rm i}^{\Pi}\rightarrow$J$_{\rm f}^{\Pi}$} & 
\multicolumn{1}{c}{10$^{-5}$}&
\multicolumn{1}{c}{(mbarn)} &
\multicolumn{1}{c}{(mbarn)}\\
\hline
$^{34}$S&2022.9$^{\rm p}$&$1/2^+\rightarrow3/2^-$& 50.73$\pm$0.91&  26.9$\pm$1.1&\\
&2082.6$^{\rm p}$&$1/2^+\rightarrow1/2^-$& 70.34$\pm$0.98& 37.2$\pm$1.4&\\
&2796.8$^{\rm p}$&$1/2^+\rightarrow1/2^-$& 27.65$\pm$0.77&  14.6$\pm$0.7&\\
&3184.0$^{\rm p}$&$1/2^+\rightarrow3/2^-$& 31.99$\pm$0.93&  16.9$\pm$0.8&\\
&4638.2$^{\rm p}$&$1/2^+\rightarrow3/2^-$&280.82$\pm$2.53&
148.6$\pm$5.5&244.3$\pm$8.8\\ 
\cline{5-6}
&1572.4$^{\rm s}$&$1/2^+\rightarrow3/2^+$&179.76$\pm$1.43&&\\
&2347.8$^{\rm s}$&$3/2^-\rightarrow3/2^+$&225.76$\pm$158&&\\
&3802.0$^{\rm s}$&$3/2^-\rightarrow3/2^+$&14.43$\pm$0.79&&\\
&4189.3$^{\rm s}$&$1/2^-\rightarrow3/2^+$&15.37$\pm$0.88&&\\
&4903.4$^{\rm s}$&$1/2^-\rightarrow3/2^+$&23.49$\pm$1.03&&\\
&4963.1$^{\rm s}$&$3/2^-\rightarrow3/2^+$&12.19$\pm$0.80&&249.3$\pm$9.0\\
\cline{5-6}
&&&& average &247$\pm$9\\
\hline
$^{36}$S&1665.7$^{\rm p}$&$1/2^+\rightarrow1/2^-$& 6.67$\pm$0.77& 54.8$\pm$6.9&\\
&2311.6$^{\rm p}$&$1/2^+\rightarrow3/2^-$& 3.81$\pm$0.59& 31.3$\pm$5.1&\\
&3657.3$^{\rm p}$&$1/2^+\rightarrow3/2^-$&18.66$\pm$0.80&
153.3$\pm$9.9&239$\pm$15\\ 
\cline{5-6}
&646.2$^{\rm s}$&3/2$^-\rightarrow7/2^-$&30.37$\pm$0.97&&249$\pm$14\\
&3103.3         &      decay line       &28.77$\pm$0.95&&236$\pm$14\\
\cline{5-6}
&&&& average &242$\pm$12\\
\hline        
\end{tabular*}
{\footnotesize $^{\rm a}$ Ref.~\cite{end90}, $^{\rm p}$ primary transition,
$^{\rm s}$ secondary ground state transition
}
\end{table}
dominates and measurements of capture cross
sections at thermal energies can be used to extract
spectroscopic 
factors for the calculation of DC cross sections
at thermonuclear energies.
\begin{figure}[htb]
\psfig{file=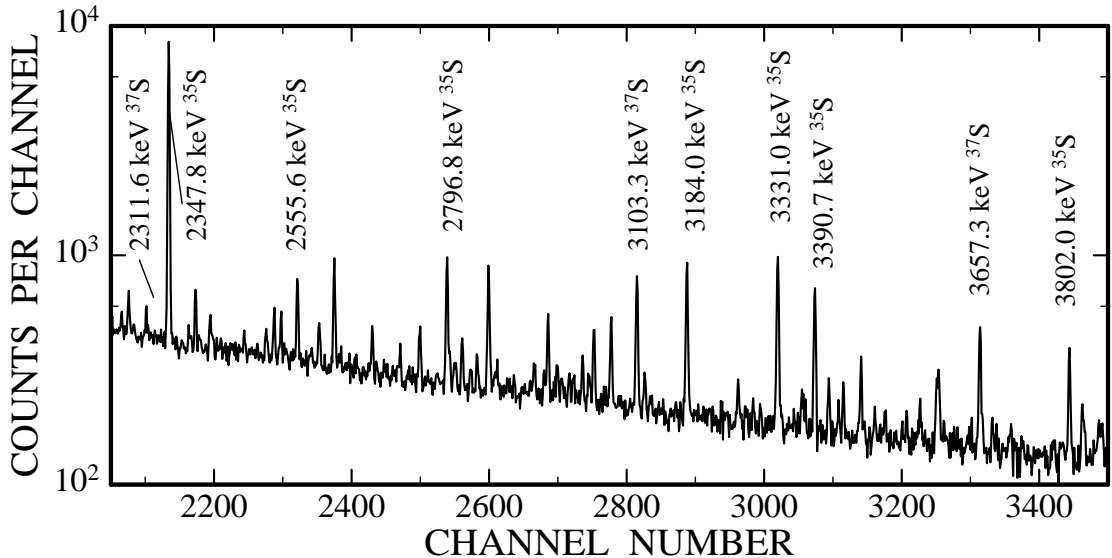,width=\textwidth}
\caption{Intensities of $^{35}$S and $^{37}$S $\gamma$--lines}
\label{fig:1}
\end{figure}
\section{EXPERIMENT}
The measurements were carried out at the reactor BR1 of the "Studiecentrum
voor kernenergie" in Mol, Belgium. The setup was the same
as in a previous experiment \cite{bee96}.
The sample of elemental sulfur (total weight 313.6\,mg) contained in a  
teflon cylinder of 6\,mm inner diameter was
enriched in $^{34}$S by 93.26\,\% and in $^{36}$S by 5.933\,\%. 
Fig.~\ref{fig:1} shows part of the
$\gamma$--ray spectrum accumulated with a Ge--detector.                 
Neutron flux and efficiency of the detector 
were determined in a calibration run on $^{35}$Cl. 
The total capture cross sections of $^{34,36}$S
were calculated from the sum of primary and secondary ground state
transitions and the $^{37}$S $\gamma$--decay line (Table \ref{t2a}).
The data were also corrected for multiple neutron scattering
($^{34}$S: 5.6\,\%, $^{36}$S: 7\,\%).
For the $^{34,36}$S cross sections systematic
uncertainties of 3.6 and 4.8\,\%, respectively, were estimated.
Our thermal $^{34}$S capture cross section is in good agreement
with the recommended value \cite{gry87}. Our $^{36}$S result is
consistent with the previous measurement of Raman et al.~\cite{ram85}.

\begin{table}[htb]
\caption{Spin/parity assignments $J^\Pi$, excitation energies $E_{\rm x}$,
and neutron spectroscopic factors of states in $^{35}$S and $^{37}$S.
The spectroscopic factors $S$(d,p)
and $S$(n,$\gamma$) are extracted
from experimental (d,p)--reactions and from the thermal neutron cross sections
measured in this work,
respectively. The calculated $\sigma^{\rm cal}$
and experimental $\sigma^{\rm exp}$ thermal capture cross sections are
compared for $^{34}$S(n,$\gamma$)$^{35}$S and $^{36}$S(n,$\gamma$)$^{37}$S.
The sum of the contributions
for all other transitions not shown in this table
give less than 1\,$\mu$barn.}
\label{t3}
\begin{tabular*}{\textwidth}{@{}l@{\extracolsep{\fill}}rrrrrr}
\hline
\multicolumn{1}{c}{$J^\Pi$} &
\multicolumn{1}{c}{$E_{\rm x}$ (MeV)} &
\multicolumn{1}{c}{$S$(d,p)} &
\multicolumn{1}{c}{$S$(n,$\gamma$)} &
\multicolumn{1}{c}{$\sigma^{\rm cal}$ (mbarn)} &
\multicolumn{1}{c}{$\sigma^{\rm exp}$ (mbarn)}\\
\hline
\multicolumn{6}{c}{$^{34}$S(n,$\gamma$)$^{35}$S}\\
\hline
3/2$^-$ & 2.348 & 0.32--0.55 & 0.48 &  99--169 & 149\\
3/2$^-$ & 3.802 &       0.09 & 0.08 &       20 &   17\\
1/2$^-$ & 4.189 & 0.12--0.14 & 0.15 &   12--14 &   15\\
1/2$^-$ & 4.903 & 0.44--0.80 & 0.49 &   33--60 &   37\\
3/2$^-$ & 4.963 & 0.18--0.22 & 0.19 &   26--32 &   27\\
\hline
&&& sum & 190--295 & 245\\
\hline
\multicolumn{6}{c}{$^{36}$S(n,$\gamma$)$^{37}$S}\\
\hline
3/2$^-$ & 0.647 & 0.44--0.70 & 0.54 & 126--200 & 153\\
3/2$^-$ & 1.992 & 0.04--0.08 & 0.19 &    7--13 &  31\\
1/2$^-$ & 2.638 & 0.48--0.80 & 0.92 &   29--48 &  55\\
\hline
&&& sum & 162--261 & 239\\
\hline
\end{tabular*}
\end{table}

\section{DIRECT--CAPTURE CALCULATIONS}

The DC formalism including the folding procedure
used in this work is described in Ref.~\cite{kra96}.
The DC calculations are performed in the same manner as given
already for $^{36}$S(n,$\gamma$)$^{37}$S in~\cite{bee95}. The
input data for the folding potentials are:
experimental charge distributions~\cite{deV87}, thermal
elastic scattering data~\cite{sea92}, excitation energies
of the final states~\cite{end90}. For the DC--calculations
the input data for the masses and Q--values are taken
from Ref.~\cite{aud93}. The spectroscopic factors are
listed in Table~\ref{t3} and are taken from (d,p)--reactions~\cite{end90}
(except the (d,p)--data from~\cite{eck89} which has
been taken from the original work, see Note added in Proof
in~\cite{bee95}).
The resulting thermal cross sections calculated in the DC--model are
compared with
the experimental cross sections for $^{34}$S(n,$\gamma$)$^{35}$S
and $^{36}$S(n,$\gamma$)$^{37}$S in the last two columns of
Table~\ref{t3}. 

\section{DISCUSSION}

The DC--mechanism reproduces not only
the summed thermal experimental cross sections
of $^{34}$S(n,$\gamma$)$^{35}$S and $^{36}$S(n,$\gamma$)$^{37}$S,
but also the transitions to the different relevant final states.
This is certainly not true for statistical model
calculations~\cite{wal81,thi87,cow91}, because
the statistical approach is valid only in the case
of a high--level density in the populated excitation range
of compound nuclei. Hauser--Feshbach
calculations for the above
reactions lead to a significant overestimation
of the cross sections by about one order of magnitude~\cite{rau96}.

The spectroscopic factors extracted
from (d,p)--reactions by different groups lead to calculated
thermal cross sections that are uncertain
by at least $\pm$\,20\,\% (see last but one column
in Table~\ref{t3}). 
Spectroscopic factors extracted
from thermal experimental (n,$\gamma$)--reactions can be determined
more accurately, because the optical potentials are much better known in
this case. Therefore, thermal (n,$\gamma$)--measurements are an
excellent alternative to (p,d)-- or other
transfer reaction measurements in determining spectroscopic
factors. These are necessary for calculating
the non--resonant direct part of the cross section
at the astrophysically relevant thermonuclear energies.

\end{document}